\documentclass{SCGE}

\usepackage{multicol}
\usepackage{hyperref}
\usepackage{graphicx}

\begin{document}

\begin{picture}(0,0){\rm
\put(0,-39){\makebox[160truemm][l]{\bf {\sanhao\raisebox{2pt}{.}}
Research Paper  {\sanhao\raisebox{1.5pt}{.}}}}}
\end{picture}

\def\bm{\boldsymbol}

\def\dl{\displaystyle}
\def\du{\end{document}}

\Year{2012} %
\Month{June}
\Vol{53} %
\No{3} %
\BeginPage{586} %
\EndPage{590} %
\AuthorMark{{\rm XIE MingJie,} et al.}
\DOI{10.1007/s11433-010-0152-8} 

\title[Multiple QPOs of NRAO~530]{Multiple periodic oscillations in the radio light curves of NRAO~530}

\author[1]{XIE MingJie}{}
\author[1]{WANG JunYi \footnote{email: wangjy@guet.edu.cn} }{}
\author[2,3]{AN Tao \footnote{email: antao@shao.ac.cn} }{}
\author[1]{ZHENG Lin}{}
\author[1]{HAN Xu}{}

\address[{\rm1}]{Key Laboratory of Cognitive Radio $\&$ Information Processing, the Ministry of Education, Guilin University of Electronic Technology, Guilin 541004, China;}
\address[{\rm2}]{Shanghai Astronomical Observatory, Chinese Academy of Sciences, Shanghai 200030, China;}
\address[{\rm3}]{Key Laboratory of Radio Astronomy, Chinese Academy of Sciences, Nanjing 210008, China}
\maketitle \vspace{-3.5mm}{\footnotesize\begin{center} Received May 21, 2012; accepted ???, 2012
\end{center}}\vspace*{-5mm}

\begin{center}
\rule{16.5cm}{0.8pt}
\parbox{16.5cm}
{\begin{abstract}
In this paper, the time series analysis method CLEANest is employed to search for characteristic periodicities in the radio light curves of the blazar NRAO~530 at 4.8, 8.0 and 14.5 GHz over a time baseline of three decades. Two prominent periodicities on time scales of $\sim$6.3 and $\sim$9.5 yr are identified at all three frequencies, in agreement with previous results derived from different numerical techniques, confirming the multiplicity of the periodicities in NRAO~530. In addition to these two significant periods, there is also evidence of shorter-timescale periodicities of $\sim$5.0 yr, $\sim$4.2 yr, $\sim$3.4 yr and $\sim$2.8 yr showing lower amplitude in the periodograms. The physical mechanisms responsible for the radio quasi-periodic oscillations and the multiplicity of the periods are discussed.
\end{abstract} }
\end{center}\vspace*{-0.6cm}

\begin{center}
\parbox{16.5cm}
{\bf\jiuhao blazar, NRAO 530, periodicity, variability, DCDFT, CLEANest}
\end{center}

\begin{center}
\parbox{16.5cm}{
\rule{16.5cm}{0.8pt}}
\end{center}

\wuhao\vspace*{1.5mm}
\begin{multicols}{2}
\renewcommand{\baselinestretch}{1.08}\selectfont
\baselineskip 12.2pt
\parindent=10.8pt


\section{Introduction}
\no The energy production and release of active galactic nuclei (AGN), which represent a population of the most powerful sources in the Universe, is one of the hottest problems in extragalactic astrophysical studies. The vast energy supply of the AGN is thought to be related to the accretion of the central supermassive black holes (SMBHs). Imaging of the energy engine of the AGN, {\it i.e.}, the BH-accretion disk system, which is extremely compact, is still limited by the angular resolution of the current observing facilities. Due to the difficulties of the imaging observations, non-imaging techniques are extensively used for investigating the dynamic processes taken place in the galactic nucleus. For instance, the flux density variability of AGN, which is rather easy to acquire from monitoring observations with a single telescope, provides vital information to constrain the energetic processes related to the black hole system. Moreover the time series analysis of the variability data is done in time or frequency domain, and thus it is free of angular resolution limitation. Blazars consist of a sub-class of AGN and are characterized by rapid and violent variability across the whole electromagnetic spectrum. Therefore variability studies of blazars open a window to understanding the AGN physics. Searching for periodicity from the variability of blazars is of particular interest, because the inferred time scale may serve to constrain the size and location of emitting source and the dynamic parameters of the physical models of AGN (e.g., [1]). Variability analysis of blazars from radio, optical, X-ray bands on diverse time scales has been reported in literature (reviewed by [2-6]).

Many numerical methods that were originally developed in statistics, signal processing and communication sciences have been applied to study the periodic oscillations in astronomical light curves. A basic tool of periodicity analysis is the Discrete Fourier Transform (DFT). However, the ordinary Fourier transform may often induce fake peaks in the power spectrum when unevenly sampled data dealt with [7-10]. Due to various observing problems, light curves of astronomical objects can be regarded as irregularly sampled time series. When the astronomical time series have large gaps, that often results in sharp spikes in the periodograms. The noise of the sparsely sampled data may also result in fake peaks in periodograms. In addition, the observed data may contain multiple modes of periodic signals, which interact with each other and make the periodogram even more complex.  An algorithm named CLEANest [11] has been designed on the basis of date-compensated discrete Fourier transform (DCDFT: [12]), and it is proven to be a robust and efficient technique to identify the real periodicities from noisy data containing multiple periodic signals (e.g., [11, 13]). The DCDFT and CLEANest method have been used for period detection of blazars [14-19] from the optical and radio light curves. The derived periods range from 2.2 to 20.8 years [15].

\begin{figure}[H]
\centering
\includegraphics[scale=0.41]{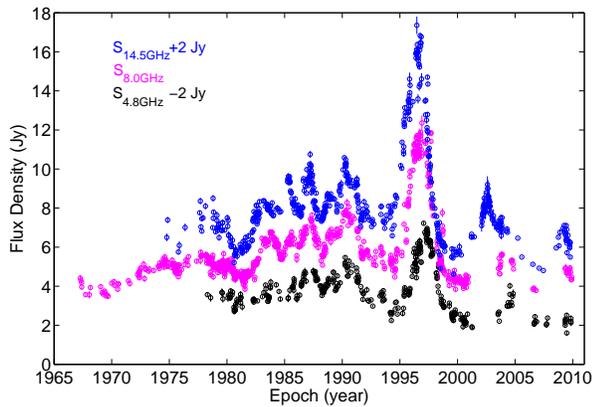}\vspace{-2mm}
\caption{\footnotesize The radio light curves of NRAO 530 at 4.8 (square), 8.0 (circle) and 14.5 GHz (triangle). In order to avoid the overlapping of three light curves, the 4.8-GHz flux densities have been subtracted by 2 Jy, and the 14.5-GHz flux densities have been added with 2 Jy. The strongest outburst occurs during 1996-1997.}
\label{fig1}
\end{figure}

NRAO 530 (PKS 1730-130) is a well-known blazar at a redshift of 0.902 [20]. The light curves of NRAO 530 show prominent outbursts across the whole electromagnetic spectrum, and VLBI observations reveal apparent superluminal motions of the jet knots, indicative of relativistic beaming of the jet flow ([21] and references therein). The radio light curves display a number of outbursts, and the peaks show an apparent 6-yr cycle since 1980 (Fig. \ref{fig1}). Previous variability analysis of the radio data led to detection of two periodicities of 10-yr and 6-yr (DCDFT: [15]; phase dispersion minimization: [22]). Except for these two most prominent periods, there were also hints of short-timescale periods [22], however the amplitudes of which were too low to be distinguished from the noise. These shorter periods need further investigation to verify their reliability. Another interesting finding is the observations of multiple periods in NRAO 530. Although multiple periods have already been observed in a number of blazars [15, 19], it is still a puzzle whether the multi-period is a common feature of the blazar variability, and the physics of the multiplicity is not well understood yet. {\bf The blazar NRAO 530 deserves a comprehensive variability study owing to the above reasons and serves as a template to better understand blazar physics.} In this paper, we shall study the periodic variability of NRAO 530 by using the monitoring data observed with the 26-meter paraboloid radio telescope of the University of Michigan Radio Observatory (UMRAO).
The observations were made at three frequencies, {\it i.e.} 14.5, 8.0 and 4.8 GHz. The light curves are exhibited in Figure \ref{fig1}.
The starting epoch of the monitoring observation  is April 1, 1978 (4.8 GHz), April 19, 1967 (8.0 GHz) and October 10, 1974 (14.5 GHz). And the latest epoch is in November 2009 for all three frequencies. The total time span is 31.6 (4.8 GHz), 42.7 (8.0 GHz) and 35.0 yr (14.5 GHz), which is much longer than the radio data used in [15] by 50\%, 33\% and 46\%, respectively. Accordingly the data points are more than [15] by 22\% (4.8 GHz), 10\% (8.0 GHz), and 16\% (14.5 GHz).
The observations before 2001 were regularly performed; an average gap between the sampling is about 2 weeks.
After 2001, there were two big gaps appearing in 2001--2003 and 2005--2008.
Considering that the typical periods of the radio variability of blazars are of the order of a few years and the strongest outburst lasts a few months, the biweekly sampling is good enough for performing an accurate periodicity analysis.

{\bf This paper will make use of the CLEANest method to search for periodicities from the radio light curves of NRAO 530. Benefitting from longer time series and more data point, the present work aims to detect possible weaker period(s) and to confirm the multiplicity of the periodicities.}
The structure of the paper is organized as follows.
In Section 2 the CLEANest algorithm is briefly described. Section 3 presents the results from the CLEANest method, and interpretations of the observations are given in Section 4.

\vspace{-2mm}
\section{The CLEANest technique}
\vspace{-2mm}

As mentioned in Section 1, classical Fourier transform of astronomical data which are irregularly sampled may often induce spurious peaks in the power spectrum due to large sampling gaps, confusing the identification of real periods. These fake peaks appear at frequencies $\omega_s \pm k\omega_g$, where $\omega_s$ is the signal frequency, $\omega_g$ is the frequency of the gaps, and $k$ is an integer [11]. The noise of the observing data may also generate fake peaks. Therefore discrimination of real periodic signals from noise is the primary task of period detection. Light curves of a substantial blazars contain multiple periods, which are interacting and blending in a complex matter. In addition, some weak short-lived periodicities might be immersed in other strong persistent periodicities. Therefore a further consideration is to effectively extract individual periodic components. Although there are many existing numerical techniques that can be used for periodicity analysis of time series data,
the CLEANest is found to be one of the excellent techniques which is powerful not only at identifying noise-induced signals, but also in separating multiple components.

\subsection{The principle of the CLEANest}

\begin{figure}[H]
\centering
\includegraphics[scale=0.40]{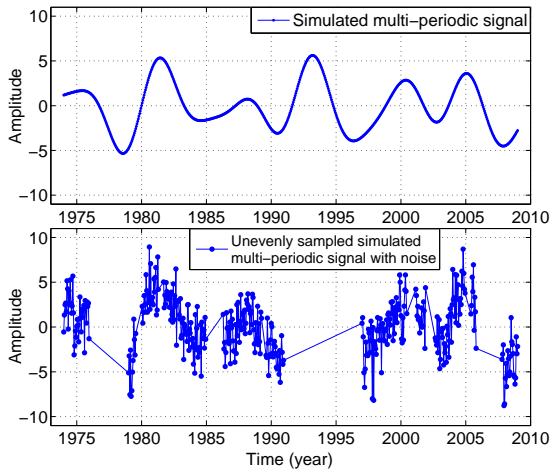}\vspace{-2mm}
\caption{\footnotesize The simulated time series mixed with three periodic components of 10, 6, and 4 yr. The top panel shows the multi-periodic signal without noise and gaps, and the bottom panel shows the signal with noises and irregular gaps added.}
\label{fig2}
\end{figure}

For periodicity analysis of unevenly spaced data, a widely used analysis technique is the modified periodogram [9, 23], which is based on a least squares regression onto the two trial functions (sin($\omega t$), cos($\omega t$)). The DCDFT is an improved Fourier-based periodogram, and is superior for dealing with the irregular sampling problems compared with ordinary Discrete Fourier Transform. It carries out least square regression fitting of the observing data with a linear combination of three trial functions, $\sin(\omega t)$, $\cos(\omega t)$ and a constant. The DCDFT improves the estimate of both the signal frequency and amplitude.

The CLEANest algorithm is exploited with an attempt to eliminate the fake peaks in periodograms resulting from irregularity of the time series.
It is in principle a combination of the DCDFT algorithm [12, 11] and the CLEAN algorithm [24, 25].
An accurate identification of the signal frequency and the following-up subtraction of the Fourier component(s) from the spectrum is achieved by utilizing the CLEAN technique [24].
In the CLEANest procedure, the strongest signal peak together with the corresponding sidelobes are first subtracted from the original DCDFT spectrum, and a residual spectrum is produced. Then the secondary component is searched for in the residual spectrum.
If the remaining strongest peak in the residual spectrum is statistically significant, the original data are analyzed again to find the pair of frequencies which give best fits to the data.
These two Fourier components (primary and secondary) and corresponding fake components (associated with their sidelobes) are subtracted again, producing a secondary-level residual spectrum.
This DCDFT-CLEAN cycle repeats until the residual spectrum is considerably flat and does not show any significant peaks.
Since the frequency components involved would be refined at each run of the CLEAN procedure, under most circumstances higher-order CLEAN
spectra would give progressively better frequency estimates.
The power spectrum obtained in this way is called the CLEANest periodogram.
The discrete CLEANed Fourier components are used to model the original dataset and create a reconstructed light curve which can be compared with the observed one.

\begin{figure}[H]
\centering
\includegraphics[scale=0.42]{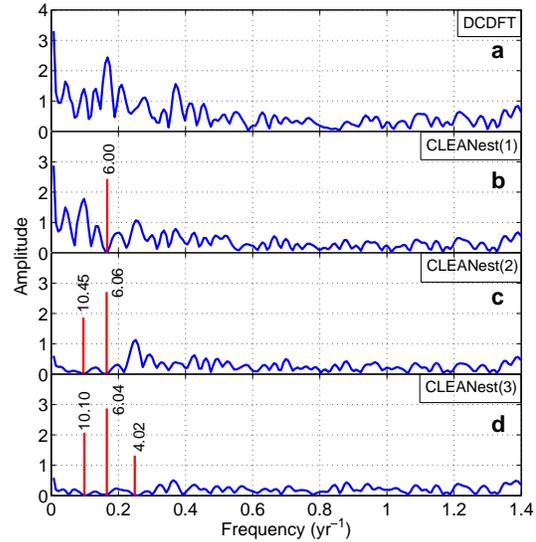}\vspace{-2mm}
\caption{\footnotesize The CLEANest periodograms of the multi-periodic simulated signal (shown in Fig. \ref{fig2} bottom panel). From top to bottom, it displays the DCDFT, the first-iteration CLEAN, the second-iteration CLEAN and the final-iteration ({\it i.e.}, the CLEANest)  spectra, in sequence. Obviously, the CLEANest(3) successfully extracts all trial periodic components.}
\label{fig3}
\end{figure}

\subsection{Testing the CLEANest algorithm with a simulated multi-periodic time series}

\no In order to test the ability of the CLEANest in extracting the underlying periodic signals from the unevenly sampled time series, we applied the CLEANest method to the simulated multi-periodic signals of $y=\sin(\frac{1}{2}\pi t)+2\sin(\frac{1}{5}\pi t)+3\sin(\frac{1}{3}\pi t)$, a combination of three sinusoidal functions with periods of 4, 10, 6 yr. These three Fourier components have magnitude-less amplitudes of 1, 2, 3 respectively. The simulated light curve is shown in the top panel of Figure \ref{fig2}. The time span of the simulated time series is from 1974 to 2009, similar to that of NRAO 530. Standard normal distribution noise and random large time gaps were also added to the simulated signals, as shown in the bottom panel of Figure \ref{fig2}.

The DCDFT scanned the whole light curve and modeled the time series with two sinusoidal functions and a constant. The derived periodogram is shown in Figure \ref{fig3}.
The highest peak is located around a frequency of 0.167 yr $^{-1}$ with an amplitude of $\sim$2.5, in rough agreement with the strongest (amplitude = 3) trial period of 6.0 yr.
To the left and right of the primary peak, there are two lower-amplitude (about 1.5) peaks that are symmetrically distributed with respect to the primary peak. These two peaks are actually the sidelobes of the strongest peak.
The secondary peak in the DCDFT is at $\sim$0.35 yr$^{-1}$ with an amplitude of $\sim$1.65.
We run an iteration of CLEAN to subtract the primary peak from the original spectrum, and obtain the residual spectrum which is shown in Figure \ref{fig3}-b.
The most significant component is at a frequency of 0.0957 yr$^{-1}$ and has an amplitude of 1.8. This Fourier component corresponds to the secondary strongest trial period of 10 yr.
The amplitude of the peak at 0.35 yr$^{-1}$ in Figure \ref{fig3}-a is significantly reduced, suggesting that it is a high-order harmonic component of the strongest periodic component generated by the Fourier transform. Therefore it is not identified as a real periodic component.
Moreover the sidelobes of the 0.167-yr$^{-1}$ primary component are also suppressed down to an amplitude level of 1.0.
Repeating the CLEAN process, we successively obtain periodograms of CLEANest(2) and CLEANest(3).
The residual spectrum of the CLEANest(3) becomes uniformly flat, having no significant spikes.
CLEAN procedure stops at this step, and all three trial periods are distinctively identified.
Note that the small shift of the detected periods away from the assumed ones reflects the intrinsic shortcoming of Fourier Transform when irregular noise is dealt with.
We can see, after each run of the CLEAN procedure, the fake high-order harmonic component and the sidelobes of the real signal are effectively extracted from the original spectrum, and they would have no much effect in the residual spectrum.

\section{Period detection in NRAO 530}

\begin{figure}[H]
\centering
\includegraphics[scale=0.40]{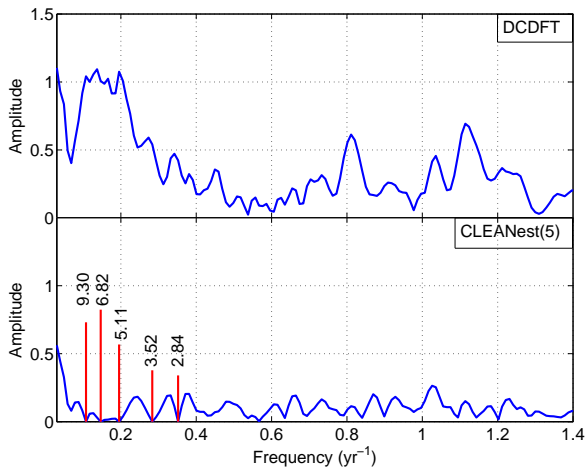}\vspace{-2mm}
\caption{\footnotesize The DCDFT (top) and CLEANest (bottom) periodograms derived from 4.8 GHz data. The DCDFT spectrum shows a broad frequency peak from 0.1075 to 0.1955 \rm yr$^{-1}$, indicating the mixture of multiple periods. The thick vertical lines in the  CLEANest periodogram (bottom) mark the subtracted periodic components.}
\label{fig4}
\end{figure}

\begin{figure}[H]
\centering
\includegraphics[scale=0.40]{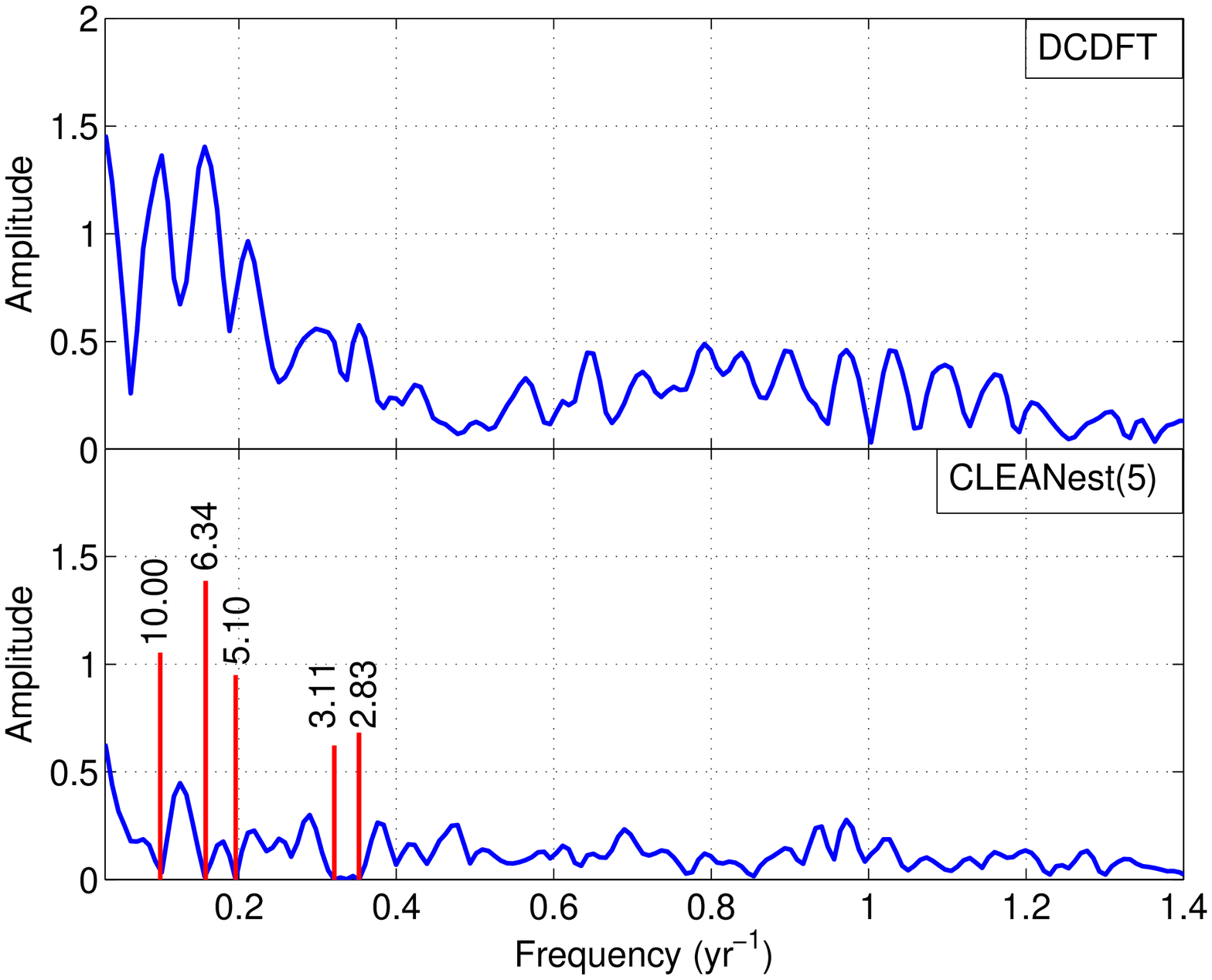}\vspace{-2mm}
\caption{\footnotesize The DCDFT (top) and CLEANest (bottom) periodograms derived for the 8.0 GHz data. }
\label{fig5}
\end{figure}

\begin{figure}[H]
\centering
\includegraphics[scale=0.40]{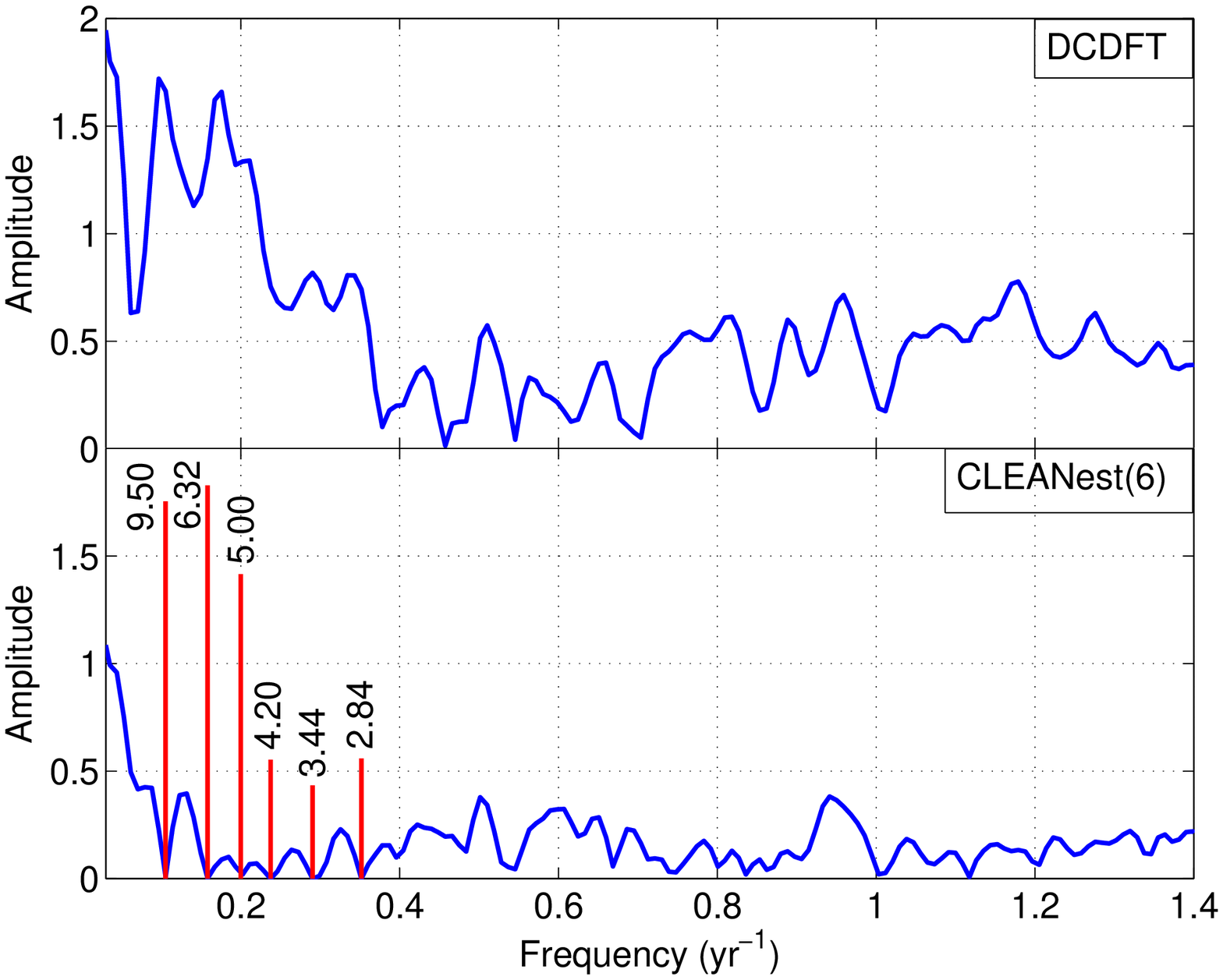}\vspace{-2mm}
\caption{\footnotesize The DCDFT (top) and CLEANest (bottom) periodograms derived for the 14.5 GHz data.}
\label{fig6}
\end{figure}

In Section 2 we have demonstrated the powerfulness of the CLEANest in fighting against the sidelobes and fake high-order harmonics. In this section, we would employ the CLEANest algorithm to search for periodicities from the radio light curves of NRAO 530.
By following the procedure described in Section 2, the DCDFT spectra were first derived from the observing data. They are displayed in the top panel of Figures \ref{fig4} to \ref{fig6}.
The DCDFT spectra show two peaks with equal amplitudes at 8.0 and 14.5 GHz (Figs. \ref{fig5} and \ref{fig6}), corresponding to frequencies of $\sim$0.1052 yr$^{-1}$ and $\sim$0.1582 yr$^{-1}$. The peaks in the 4.8-GHz DCDFT spectrum (Fig. \ref{fig4}) are distributed in a rather broad range from frequency 0.1075 to 0.1955 \rm yr$^{-1}$, indicating a mixture of multiple Fourier components with close frequencies. There are also two high-frequency peaks appearing at 0.8112 yr$^{-1}$ and 1.1140 yr$^{-1}$ in Fig. \ref{fig4}, respectively.
The two strongest peaks at frequencies of 0.1052 yr$^{-1}$ (period = 9.53 yr) and 0.1582 yr$^{-1}$ (period = 6.32 yr) have already been detected in previous research with the same DCDFT method [15], and later confirmed with the PDM method [22]. Therefore we used these two periods which are confirmed as initial trial periodic components and first subtracted them from the DCDFT spectra. In the resulting 4.8-GHz residual spectrum, a Fourier component at frequency $\sim$0.2000 yr$^{-1}$ became the strongest. The two peaks shown at 0.8112 yr$^{-1}$ and 1.1140 yr$^{-1}$ in the DCDFT spectrum became less significant, suggesting they are probably the higher-order harmonics of the 9.53-yr and/or 6.32-yr periodic components. The 5-yr component was also detected in CLEANest spectra at 8.0 and 14.5 GHz. Fan et al. [15] only detected this 5-yr periodic component from the DCDFT spectrum at 4.8-GHz, whereas this component appears at all three frequencies in our CLEANest periodograms. The derived
amplitude of the 5-yr periodic component is comparable to that of the 10-yr component. Thus we regarded it as a possible period and subtracted it from the spectrum. After subtracting the three strong periodic components, the remaining peaks in the residual spectra were lower than half of the maximum of the original DCDFT spectrum. We further identified two components from the 4.8-GHz spectrum, at frequencies of 0.2834 yr$^{-1}$, and 0.3518 yr$^{-1}$, corresponding to periods of 3.52 and 2.84 yr, respectively. The 8.0-GHz CLEANest periodogram also presented the 3.11-yr and 2.83-yr periodic components. The peaks in the residual spectra (bottom panels in Figures \ref{fig4} and \ref{fig5}) became less convincing, and then the CLEANest stopped. The same procedure was applied to the 14.5-GHz data, and we got three periodic components at about 4.20, 3.44 and 2.84 yr.
The CLEANed periodic components are tabulated in Table \ref{tab1}.

\begin{figure}[H]
\centering
\includegraphics[scale=0.42]{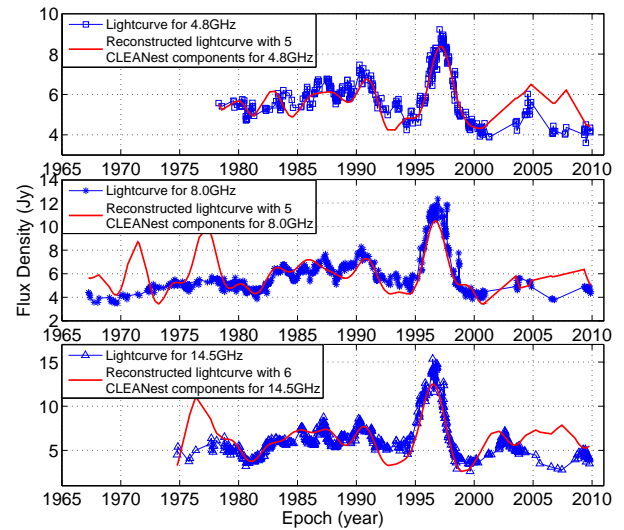}\vspace{-2mm}
\caption{\footnotesize The reconstructed light curves (thick lines) overlaid on observed light curves. }
\label{fig7}
\end{figure}

In order to verify the detected periodicities, the CLEANed Fourier components were used to model the original datasets and generate reconstructed light curves of NRAO 530 which were compared with the observed light curves in Figure \ref{fig7}.
The reliability of the detected periodicities is manifested by the goodness of the fits of two sets of light curves.
At all three frequencies, the reconstructed light curves basically recover the observed ones from epochs 1980 to 2000.
The fits later than epoch 2000 and earlier than 1980 show large deviations because the data points are sparsely sampled in these time spans.
Alternatively the deviation between the modeled and observed light curves arises because the CLEANed components are fitted with constant amplitudes. However, the apparent magnitudes of the periodic radio outbursts may alter with time considering the complex instabilities of the jet flow and/or possibly varying Doppler boosting factors.
We also tried using different numbers of the periodic components, and found that the model light curves constructed using all five or six CLEANed frequency components give best match with the observational light curves, recovering more fine structures in the light curves.

\begin{table}[H]
\begin{center}
\caption{\footnotesize The CLEANest frequency components of NRAO 530}
\label{tab1}
\vspace{1mm}
\begin{tabular}{cccccl}
 \hline\hline
$\nu_{obs}$ (GHz) &  \#  & Freq. (yr$^{-1}$) & P (yr)  & Amp.  \\
\hline
    & 1 & 0.105263158     & 9.50  & 1.7536  \\
    & 2 & 0.158227848     & 6.32  & 1.8280  \\
14.5& 3 & 0.200000000     & 5.00  & 1.4151  \\
    & 4 & 0.237572585     & 4.20  & 0.5523  \\
    & 5 & 0.290366492     & 3.44  & 0.4338  \\
    & 6 & 0.351959385     & 2.84  & 0.5585  \\ \hline
    & 1 & 0.100000000     &10.00  & 1.0530  \\
    & 2 & 0.157728707     & 6.34  & 1.3864  \\
8.0 & 3 & 0.195936372     & 5.10  & 0.9490  \\
    & 5 & 0.321335651     & 3.11  & 0.6820  \\
    & 6 & 0.352685470     & 2.83  & 0.6220  \\ \hline
    & 1 & 0.107507666     & 9.30  & 0.7308  \\
    & 2 & 0.146601363     & 6.82  & 0.7814  \\
4.8 & 3 & 0.195468484     & 5.11  & 0.5677  \\
    & 5 & 0.283429302     & 3.52  & 0.3781  \\
    & 6 & 0.351843271     & 2.84  & 0.3405  \\ \hline
   \hline\hline
\end{tabular}
\end{center}
\end{table}

In summary, all together five or six possible periods have been detected from the NRAO 530 radio light curves by using the CLEANest approach. The two strongest periods of $\sim$9.5-yr and $\sim$6.3-yr have also been reported by other authors using the DCDFT and PDM techniques [15, 22]. A 5.0-yr periodicity is detected in the CLEANest periodograms at all three observing frequencies, identified as a possible period.
In addition, there are three possible periods which are newly found in the present work.
They appear at a higher frequency regime, corresponding to characteristic periods of 4.2, 3.4, and 2.8 yr.
A possible reason for the new detection of three higher-frequency periodic components is that the current analysis uses more data points across a longer time span, which results in lower noise in the residual spectra. The relatively weaker peaks have more chance to be detected.
Moreover these weaker periodic components might not be persistent across the whole time span.
When DCDFT carries out a global modeling of the light curves with a linear combination of trial functions, any short-lived or intermittent periodic components would probably be smoothed in the global power spectrum.
In the previous search for periodicity in NRAO 530 [15] the weaker periodic components are probably hidden among the relatively higher noise due to shorter time coverage and less data points in their study.

\section{Discussion and Conclusion}

A robust Fourier-based algorithm CLEANest was used to search for periodicities from the radio light curves of the blazar NRAO 530. Five periodic components were detected at 4.8 and 8.0 GHz, and six components at 14.5 GHz. The basic properties of the detected periodicities can be summarized as :
\begin{itemize}
 \item multiplicity of the periodicities.
The present work confirms that the radio flux densities of NRAO 530 show periodic oscillations and there are multiple underlying periods. Two periods on time scale of 9.5 and 6.3 yr are most prominent, and they together dominate the power spectrum. These two periodicities have been identified at all frequency bands and with various methods. Monte Carlo tests of the statistical confidence of these two periods gave a high confidence level (Lu et al. 2012). In addition, there are four other weaker periodic components, 5.1 yr, 4.2 yr, 3.4 yr and 2.8 yr. Quantitative statistical confidence estimates of these weaker and shorter periods are necessary to clarify their reliability. A possible reason for the weakness of these shorter periods is that they are associated with short-lived or intermittent periodic components. Their integrated powers over the whole time span are relatively lower than those of persistent periodic components. Fourier transform carries out a global modeling of the entire time series, which tends to wash out the (sharp) peaks of short-lived components in the power spectrum. These newly detected shorter periods can not be the high-order harmonics of the strongest 9.5-yr and 6.3-yr periods: in each iteration of CLEAN, the periodic components identified from the previous iteration would be subtracted from the DCDFT spectrum, while at the same time, the sidelobes and high-order harmonic frequency components are removed as well. This effect is nicely manifested in the tests of simulated time series in Section 2 and Figure \ref{fig2}.

\item mixture of persistent and short-lived periodicities.
As discussed above, short-lived periodic components might be submerged by dominant persistent components. Therefore it would be necessary to seek more sophisticated algorithms which have high resolution in both time and frequency domains to identify the transient periodicities. In fact, in a complementary periodicity analysis using the wavelet transform, which is superior to Fourier-transform methods in the localization properties in both time and frequency domains, indeed reveals that the 9.5- and 6.3-yr periods are persistent across the observing session, while other periodic components only appear in certain time ranges. Mixture of multiple periodicities, and the interactions of these periodicities increase the difficulty discerning discrete individual components.

\item possible harmonic relations of the characteristic frequencies?
The present and previous work provides strong evidence for the existence of multiple periodicities in NRAO 530 (at least the strongest ones, 9.5 yr, 6.3 yr and 5.0 yr are rather significant). If all these periodicities are real, an interesting  question should be answered: do they have intrinsic relationship?  Looking at the parameters of these periodicities listed in Table 1, we find the characteristic frequencies have nearly integral ratios. From P1 to P6, the ratios are approximately 2 : 3 : 4 : 5 : 6 : 7. If the relationship of integral ratios is confirmed in the future, it would be robust evidence that these periodicities have a common origin in physical nature.
\end{itemize}

The radio emission of blazars is generally thought to arise from relativistic jets. A straightforward interpretation of periodic radio variability is attributed to the jet precession. The periodic outburst is related to periodic ejection of new jet components. In this respect, one would expect to see correlation between the jet ejection and radio outburst. The jet precession could be triggered by the orbital motion of a binary BH system [26], or a warping accretion disk which results from the Lense-Thirring torque due to Bardeen-Petterson effect ([27] and references therein). The inferred time scales in the above processes vary from hours to years, depending on the mass and spin of the black hole as well as the orbital properties (e.g., orbital period for binary BH model, or transition radius between the inner flat and outer tilted disk for warping disk model).
However the precession model encounters difficulty in the interpretation of the multiplicity and possible harmonic relationship of the detected periodicities.
Alternatively, the periodic oscillations of flux densities may result from the hydrodynamic instabilities of the jet flow. One commonly-seen instability in relativistic jets is Kelvin-Helmoholtz (K-H) instabilities [28] that can be triggered by either regular perturbations to the jet base (e.g. precession of the accretion disk), or random perturbations (e.g., jet-cloud interactions). Under certain circumstances, initial perturbations may grow and propagate along the jet flow as waves with various modes such as pinch (n$=$0), helical (n$=$1), elliptical (n$=$2). The hydrodynamic instabilities may shape the jet morphology, and also result in periodic flux variations. Lobanov \& Zensus successfully modeled the emission structure of the 3C~273 jet with a combination of helical and elliptical surface (m$=$0) and internal (m$>$1) normal modes [29]. Similar approaches were also adopted to study the M87 jet [30] and 3C 120 jet [31]. The instability model naturally explains the multiplicity and harmonic relationship of the periodicities in NRAO 530. For instance, the propagating shocks in the helically twisted jets periodically pass through the light of sight, causing the periodic enhancement of the flux density owing to Doppler boosting effect. The observed brightness distribution of the jet represents a combination of multiple instability modes. Further monitoring of the jet structure variation and temporal variability would be necessary to discriminate the different models.

\Acknowledgements{\bahao
{\bf We thank the referee very much for his/her valuable comments, which have been very helpful for improving the manuscript.}
This work has been partly supported by the 973 Program of China (2009CB24900, 2012CB821800), the Strategic Priority Research Program of the Chinese Academy of Sciences (XDA04060700), the National Natural Science Foundation of China(61261017). The University of Michigan Radio Astronomy Observatory is supported by funds from the NSF, NASA, and the University of Michigan.}


\normalsize \vskip0.3in\parskip=0mm \baselineskip 18pt
\renewcommand{\baselinestretch}{1.1}\footnotesize\parindent=4mm\bahao


\REF{1\ }Carrasco L, Dultzin-Hacyan D, Cruz-Gonzalez I. Periodicity in the BL Lac object OJ~287. Nature, 1985, 314: 146--148

\REF{2\ }Altschuler D R. Variability of extragalactic radio sources: theory and observation. Fundam Cosmic Phys, 1989, 14: 37--129

\REF{3\ }Mushotzky R F, Done C, Pounds K A. X-ray spectra and time variability of active galactic nuclei. Ann Rev Astron Astrophys, 1993, 31: 717--761

\REF{4\ }Wegner S J, Witzel A. Intraday Variability In Quasars and BL Lac Objects. Ann Rev Astron Astrophys, 1995, 33: 163--198

\REF{5\ }Ulrich M, Maraschi L, Urry C M. Variability of active galactic nuclei. Ann Rev Astron Astrophys, 1997, 35: 445--502

\REF{6\ }Abdo A A, Ackermann M, Ajello M, et al. Gamma-ray Light Curves and Variability of Bright Fermi-detected Blazars. Astrophys J, 2010, 722: 520--542

\REF{7\ }Lomb N R. Least-squares frequency analysis of unequally spaced data. Astrophys Space Sci, 1976, 39: 447--462

\REF{8\ }Scargle J D. Studies in astronomical time series analysis. I - Modeling random processes in the time domain. Astrophys J Suppl, 1981, 45: 1--71

\REF{9\ }Scargle J D. Studies astronomical time analysis. II. Statistical aspects of spectral analysis of unevenly spaced data. Astrophys J, 1982, 263: 835--853

\REF{10\ }Scargle J D. Studies in astronomical time series analysis. III. Fourier transforms, autocorrelation functions, and cross-correlation functions of unevenly spaced data. Astrophys J, 1989, 343: 874--887

\REF{11\ }Foster G. The CLEANest Fourier Spectrum. Astron J, 1995, 109: 1889--1902

\REF{12\ }Ferraz-Mello S. Estimation of periods from unequally spaced observations. Astron J, 1981, 86: 619--624

\REF{13\ }Percy J R, Wilson J B, Henry G W. Long-Term VRI Photometry of Small-Amplitude Red Variables. I. Light Curves and Periods. Publ Astron Soc Pac, 2001, 113: 983--996

\REF{14\ }Fan J H, Tao J, Qian B C, et al. Optical Photometrical Observations and Variability for Quasar 4C 29.45. Publ Astron Soc Japan, 2006, 58: 797--808

\REF{15\ }Fan J H, Liu Y, Yuan Y H, et al. Radio variability properties for radio sources. Astron Astrophys, 2007, 462: 547--552

\REF{16\ }Yuan Y H, Fan J H. Long term periodicity analysis of the spectral index of 2251+158. Research in Astron Astrophys, 2011, 11: 286--292

\REF{17\ }Yang J H, Fan J H, Liu Y, et al. Long-term Variability Properties of Mkn501. Chin Astron Astrophys, 2008, 32: 129--139

\REF{18\ }Fan J H, Liu Y, Qian B C, et al. Long-term variation time scales in OJ 287. Research in Astron Astrophys, 2010, 10: 1100--1108

\REF{19\ }Tao J, Fan J H, Qian B C, et al. Optical monitoring of 3C 390.3 from 1995 to 2004 and possible periodicities in the historical light curve. Astron J, 2008, 135: 737--746

\REF{20\ }Junkkarinen V. Spectrophotometry of the QSO NRAO 530. Publ Astron Soc Pac, 1984, 96: 539--542

\REF{21\ }Hong X Y, Sun C H, Zhao J H, et al. Bending of Jets in the QSO NRAO 530. Chin J Astron Astrophys, 2008, 8: 179--194

\REF{22\ }Lu J C, Wang J Y, An T, et al. Periodic radio variability in NRAO 530: phase dispersion minimization analysis. Research in Astron Astrophys, 2012, 12: 643--650

\REF{23\ }Horne J, Baliunas S. A prescription for period analysis of unevenly sampled time series. Astrophys J, 1986, 302: 757--763

\REF{24\ }H\"{o}gbom J A. Aperture Synthesis with a Non-Regular Distribution of Interferometer Baselines. Astron Astrophys Suppl, 1974, 15: 417--426

\REF{25\ }Roberts D H, Leh\'{a}r J, Dreher J W. Time Series Analysis with Clean. I. Derivation of a Spectrum. Astron J, 1987, 93: 968--989

\REF{26\ }Begelman M C, Blandford R D, Rees M J. Massive black hole binaries in active galactic nuclei. Nature, 1980, 287: 307--309

\REF{27\ }Lei W H, Zhang B, Gao H. Frame-dragging, disk warping, jet precessing, and dipped X-ray lightcurve of Sw J1644+57. arXiv1202.4231, 2012

\REF{28\ }Hardee P E. Modeling Helical Structures in Relativistic Jets. Astrophys J, 2003, 597: 798--808

\REF{29\ }Lobanov A P, Zensus J A. A Cosmic Double Helix in the Archetypical Quasar 3C~273. Science, 2001, 294: 128--131

\REF{30\ }Lobanov A P, Hardee P, Eilek J. Double Helix in the Kiloparsec-Scale Jet in M 87. Astron Soc Pac, 2005, 340: 104--106

\REF{31\ }G\'omez J L, Marscher A P, Alberdi A, et al. Monthly 43 GHz VLBA Polarimetric Monitoring of 3C 120 over 16 Epochs: Evidence for Trailing Shocks in a Relativistic Jet. Astrophys J, 2001, 561: L161--L164

\end{multicols}
\end{document}